\documentclass[twocolumn,showpacs,preprintnumbers,amsmath,amssymb]{revtex4}

\usepackage{epsfig}
\usepackage{graphicx}
\usepackage{dcolumn}
\usepackage{bm} 
\newcommand{\beq}{\begin{equation}}
\newcommand{\eeq}{\end{equation}}
\newcommand{\beqa}{\begin{eqnarray}}
\newcommand{\eeqa}{\end{eqnarray}}
\newcommand{\lam}{\lambda}

\newcommand{\rh}{\rho}
\newcommand{\ga}{\gamma}

\newcommand{\al}{\alpha}
\newcommand{\si}{\sigma}

\newcommand{\om}{\omega}

\newcommand{\ra}{\rangle}

\begin{document}

\title{Finite-Time Disentanglement via Spontaneous Emission}

\author{Ting Yu}
\email{ting@pas.rochester.edu}

\author{J.\ H.\ Eberly}
\email{eberly@pas.rochester.edu}
\affiliation{ Rochester Theory Center for Optical
Science and
Engineering and the Department of Physics \& Astronomy\\
University of Rochester, Rochester, New York 14627
}


\date{9 August 2004}

\begin{abstract} We show that under the influence of pure vacuum
noise two entangled qubits become completely disentangled in a finite
time, and in a specific example we find the time to be given by
$\ln \Big(\frac{2 +\sqrt 2}{2}\Big)$ times the usual spontaneous lifetime.
\end{abstract}

\pacs{03.65.Yz, 03.65.Ta, 42.50.Lc}

\maketitle

Superposition and entanglement are two basic features that
distinguish the quantum world from the classical world. While
quantum coherence is recognized as a major resource, decoherence
due to the interaction with an environment is a crucial issue that
is of fundamental interest \cite{defdeco,zeh,str,ah}. When coherence
exists among several distinct quantum subsystems the issue becomes
more
complicated because, along with the local coherence of each
constituent particle, their entanglement brings a special kind of
distributed or nonlocal coherence. It is this
distributed coherence that really matters in many important
applications of quantum information \cite{preskill,mc}.
Consequently, the fragility of nonlocal quantum coherence is
recognized as a main obstacle to realizing quantum computing and
quantum information processing (QIP) \cite{viola1,plk}. Apart from
the important link to QIP realizations, a deeper understanding of
entanglement decoherence is also expected to lead to new insights
into quantum fundamentals, particularly quantum measurement and
the quantum-classical transition \cite{gisin1,hal,dio}. Although
quantum decoherence has been extensively studied in recent years,
it remains unclear how a local decoherence rate is related to a
nonlocal disentanglement rate when a multi-particle quantum state
is in contact with one or more noisy environments.

\begin{figure}[!b]
\epsfig{file=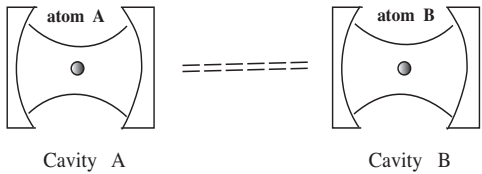, height=2.3 cm, width=6.3cm}
\caption{\label{fig} Schematic illustration of a set-up in which
two atoms are located inside two spatially separated cavities A
and B. The two atoms are initially entangled but have no direct
interaction afterwards.}
\end{figure}
Therefore, a deep understanding of the decoherence in any viable
realization of qubits is desirable and it is surprising that few if
any fundamental treatments exist of decoherence that include the
dynamics of disentanglement on better than an empirical or
phenomenological basis.

Here we consider two initially entangled
qubits and examine the dynamics of their disentanglement due to
spontaneous emission without phenomenological approximation.
There is perhaps no simpler realistic bipartite
model in which all of the effects of quantum noise can be considered
fully analytically. We show that decoherence caused by vacuum
fluctuations can affect localized and distributed coherences in very
different ways. As one surprising consequence, we show that spontaneous
disentanglement may take only a finite time to be completed,
while local decoherence (the normal single-atom transverse and
longitudinal decay) takes an infinite time.

To make our model and results concrete, we restrict our attention
to two two-level atoms $A$ and $B$ coupled individually to two
cavities which are initially in their vacuum states (see Fig.~
\ref{fig}). In the general framework of system-plus-environment,
the two two-level atoms are identified as the system of interest,
whereas the two cavities serve as the environments. The interaction
between each atom and its environment results in the loss of
both local coherence and quantum entanglement of the two atoms.
In its simplest form such a model may be formulated with
the following total Hamiltonian, which is given by (we set
$\hbar=1$): $H_{\rm tot}= H_{\rm at} + H_{\rm int} + H_{\rm cav}$,
where the Hamiltonian of the two atoms
$H_{\rm at }$, the two cavities $H_{\rm cav}$ and the
interaction $H_{\rm int}$ are given by
\beqa
H_{\rm at} &=&
\frac{1}{2}\om_A \si^A_z +\frac{1}{2}\om_B \si^B_z\label{eq1} \\
H_{\rm cav} &=& \sum_{\bm{k}}\om_{\bm k}a_{\bm k}^\dag a_{\bm k} +
\sum_{\bm k} \nu_{\bm k}b_{\bm k}^\dag b_{\bm k}\label{eq2}\\
H_{\rm int} &=& \sum_{{\bm k}} ( g^*_{\bm k}\si^A_-a^\dag_{\bm k} +
g_{\bm k} \si^A_+a_{\bm k})\nonumber\\
&+& \sum_{{\bm k}} (f^*_{\bm k}\si^B_-b^\dag_{\bm k}
+ f_{\bm k}\si^B_ + b_{\bm k})
\label{eq3}
\eeqa
where $g_{\bm k}, f_{\bm k}$ are coupling constants and
$\si_z$ denotes the usual diagonal  Pauli matrix,
and the standard 2-qubit product basis is given by:
\begin{eqnarray}\label{basis}
|1\ra_{AB}&=&|++\ra_{AB},\quad
|2\ra_{AB}=|+-\ra_{AB},\nonumber\\
|3\ra_{AB}&=&|-+\ra_{AB},\quad
      |4\ra_{AB}=|--\ra_{AB},
\end{eqnarray}
where $|\pm\pm\ra_{AB} \equiv |\pm\ra_A\otimes |\pm\ra_B$ denote the
eigenstates of the product
Pauli spin operator $\sigma_z^A\otimes\si_z^B$ with eigenvalues
$\pm 1$. The total Hamiltonian, given by equations
(\ref{eq1})-(\ref{eq3}), provides us an important solvable
model of the atom-field interaction in quantum optics.

Suppose that initially the atoms are entangled with each other but
not with the cavities, i.e., we assume that at $t=0$ the two atoms
and the cavities are described by the product state,
\beq
|\Psi_{\rm tot}\rangle = |\psi\rangle_{AB}\otimes |0\rangle_A|0\rangle_B,
\eeq
where $|\psi\rangle_{AB}$ is the entangled initial state of the two atoms
and
$|0\rangle_A|0\rangle_B$ is the vacuum state of two cavities. For
simplicity, we will not take into account the
spatial degrees of freedom of the two atoms.  The convolutionless
master equation of the system of two atoms can be obtained as
follows \cite{com0}:
\beqa
\frac{d}{dt}\rho&=&-i[H'_{\rm at}(t), \rho]\nonumber\\
&& + F_R(t)\Big([\si_-^A,\rho_t\si_+^A]+[\si_-^A\rh_t,\si_+^A]\Big)
\nonumber\\
&& + G_R(t)\Big([\si_-^B,\rho_t\si_+^B] + [\si_-^B\rh_t,\si_+^B]\Big)
\label{master}
\eeqa
where $H'_{\rm at}(t)$ is the system Hamiltonian modified to take into
account
the Lamb shifts
\beq
H'_{\rm at}(t) = \frac{1}{2}(\om_A+F_I(t))
\si^A_z +\frac{1}{2}(\om_B+G_I(t)) \si^B_z
\eeq
and  the
coefficients $F_{R,I}$ and $G_{R,I}$ are the real and imaginary
parts of $F(t)$ and $G(t)$, which are given by
\beqa
F(t) &=& \frac{1}{b(t)}\int^t_0ds\al(t-s)b(s)\\
G(t) &=& \frac{1}{c(t)}\int^t_0ds\beta(t-s)c(s)
\eeqa
Note that the cavity correlation functions are
\beqa
\al(t-s) &=& \sum_{\bm k}|g_{\bm k}|^2e^{-i\omega_{\bm k}(t-s)}\\
\beta(t-s) &=& \sum_{\bm k}|f_{\bm k}|^2 e^{-i\nu_{\bm k}(t-s)}
\eeqa
and the functions $b(s)$ and $c(s)$ are the fundamental
solutions of the equations of motion:
\beqa
\dot{b}(s) &+& i\om_A b(s) + \int^s_0 d\lambda
\al(s-\lambda)b(\lambda)=0\\
\dot{c}(s) &+& i\om_B c(s)+\int^s_0 d\lambda \beta(s-\lambda)c(\lambda)=0
\eeqa

The master equation (\ref{master}) is extremely useful for the
study of decoherence, and for the purpose of disentanglement analysis
it will be very convenient to find an explicit expression for its
solution. In the interaction picture, where
\beq
\tilde{\rho}(t) = e^{i\int^t_0 H'_{\rm at}(s)ds} \rh(t)
e^{-i\int^t_0 H'_{\rm at}(s)ds},
\eeq
the general solutions of equation (\ref{master}) can be described in
terms of a Kraus representation \cite{preskill,mc,kra,wk}. As can be
seen below, the Kraus representation allows a very elegant analysis
of the disentanglement time for arbitrary states. Precisely, for any
initial state $\rho(0)$, the density operator at $t$ can be expressed
as\cite{com}
\begin{equation} \label{sum}
\tilde{\rho}(t) =\sum_{\mu=1}^4K_{\mu}(t) \rh(0)K^\dag_\mu(t),
\end{equation}
where the Kraus operators $K_\mu(t)$ satisfy
$\sum_\mu K^\dag_\mu K_\mu=I$ for all $t$. The Kraus operators for this
model
are given by
\begin{eqnarray}
      \label{k1}K_1
&=&\left(\begin{array}{clcr}
\gamma_A && 0\\
0 && 1\\
\end{array}
       \right)\otimes  \left(
\begin{array}{clcr}
\ga_B & 0\\
0 & 1\\
\end{array}
       \right),\\
       K_2&=&\left(\begin{array}{clcr}
\gamma_A && 0\\
0 && 1\\
\end{array}
       \right)\otimes\left(
\begin{array}{clcr}
0 & 0 \\

\om_B & 0\\
\end{array}
       \right),\\
K_3&=& \left(
\begin{array}{clcr}
0 & 0\\
\om_A & 0\\
\end{array}
       \right) \otimes \left(
\begin{array}{clcr}
\ga_B & 0\\
0 & 1\\
\end{array}
       \right),\\
K_4 &=&
       \left(
\begin{array}{clcr}
0 & 0\\
\om_A & 0\\
\end{array}
       \right)\otimes \left(
\begin{array}{clcr}
0 &  0 \\
\om_B & 0\\
\end{array}
       \right),\label{k5}
       \end{eqnarray}
and the time-dependent Kraus matrix elements are
\begin{eqnarray}
\gamma_A(t)&=& \exp\left[{-\int^t_0 ds F_R(s)}\right],\label{cor1}\\
\gamma_B(t)&=& \exp\left[{-\int^t_0 ds G_R(s)}\right], \\
\om_A(t)&=&\sqrt{1-\gamma^2_A(t)},\,\,\,\,
\om_B(t)=\sqrt{1-\gamma^2_B(t)}\label{cor2}
\end{eqnarray}

With the preceding discussion, we are now in a position to determine both
the local decoherence rate and the disentanglement rate. Multiple
interpretations of the term decoherence in the literature can lead to
confusion,
so we will use global or non-local decoherence (or disentanglement)
here when we refer to loss of bipartite entanglement. The terms
local decoherence or local relaxation will refer to
longitudinal and transverse decay of single-atom density matrix elements.
In the present example local and non-local decoherence both
arise from the effects of spontaneous emission and in that sense are not
independent.

We begin with the coherence decay of a single qubit under the master
equation
(\ref{master}). The local decoherence rates of the qubits can be
estimated from the reduced density matrices
$\rho^{A}\equiv{\rm Tr}_B\{\rh\} \quad{\rm and}
\quad \rho^{B}\equiv{\rm Tr}_A\{\rh\}$. The local decoherence rates are
determined by the well-known Bloch equations with general time-dependent
functions $F(s), G(s)$. For example,
\beq
\langle \dot{\si}^A_-\rangle = -\left[F(t)+i\om_A\right]\langle
\si^A_-\rangle
\label{deph}
\eeq
and $\langle \dot{\si}^A_+\rangle =\langle \dot{\si}^A_-\rangle^*$,
where $\si^A_-=\si_-\otimes I$. Similar equations hold for
$\si_\pm^B$. Hence we have
\beqa
|\langle
\si^A_\pm(t)\rangle| &=& |\langle \si^A_\pm(0)\rangle|e^{-\int^t_0 ds
F_R(s)},\\
|\langle \si^B_\pm(t)\rangle| &=& |\langle
\si^B_\pm(0)\rangle|e^{-\int^t_0 ds G_R(s)}.
\eeqa

Given these equations, local decoherence behaviors are determined
by the character of the functions $F_R(t)$  and $G_R(t)$ which we
always assume to be positive functions asymptotically. In the familiar
Born-Markov approximation one has purely exponential decay with
rates given by $F(t)\rightarrow \Gamma_A/2$ and
$G(t)\rightarrow \Gamma_B/2$, where the $\Gamma$'s are the Einstein A
coefficients for the two-level atoms in the cavities.

The comparison of interest is with the disentanglement rate. Since
entanglement decoherence processes are most generally associated with
mixed states, we will use Wootters's concurrence to quantify the
degree of entanglement \cite{woo}. The concurrence is conveniently
defined for both pure and mixed states.  Let $\rho$ be a density
matrix of the pair of atoms expressed in the standard basis
(\ref{basis}). The concurrence may be calculated explicitly from
the density matrix $\rho$ for qubits A and B: $C(\rh) =
\max\left(0,\sqrt{\lam_1}-\sqrt{\lam_2}-\sqrt{\lam_3}
-\sqrt{\lam_4}\,\,\right), \label{defineconcurrence}$ where the
quantities $\lam_i$ are the eigenvalues of the matrix $\zeta$:
\beq
\zeta \equiv \rho(\sigma^A_y\otimes \sigma^B_y)\rho^*(\sigma^A_y\otimes
\sigma^B_y),
\label{concurrence}
\eeq
arranged in decreasing
order. Here $\rh^*$ denotes the complex conjugation of $\rh$ in
the standard basis (\ref{basis}) and $\si_y$ is the usual (pure imaginary)
Pauli matrix
expressed in the same basis. It can be shown that the concurrence
varies from $C=0$ for a
disentangled state to $C=1$ for a maximally entangled state.

We now show two categories of result for entanglement decay.
In the first more general category we show that, for all entangled
(possibly mixed) states, entanglement decays not only
more rapidly than the fastest decoherence rate of an individual
qubit, but at least as fast as the sum of the separate rates. In the
second
category we present a sharper result in a specific mixed state example,
in which the entanglement goes exactly to zero in a finite time and
remains
zero. Both categories of result are a consequence of normal
spontaneous emission.

For the first category of result, let us note that the concurrence
$C(\rh)$ is a
convex function of $\rho$ \cite{woo}. From (\ref{sum}), one immediately
has
\begin{equation}
C(\rh(t)) \leq \sum_{\mu=1}^4C(K_\mu\rh(0)
K^\dag_\mu),
\label{ineq}
\end{equation}
where $K_\mu$ are defined in equations (\ref{k1}) to (\ref{k5}).
Let us consider a typical term $C(K_\mu\rh(0)K^\dag_\mu)$ in
(\ref{ineq}) and denote it by $\rh_{\mu}=K_\mu\rh(0)K^\dag_\mu.$
  From the definition of concurrence, it can be proved that
\beqa
C(\rho_1) &=& e^{-\int^t_0(F_R(s)+G_R(s))ds}C(\rh(0)),\\
C(\rho_\mu) &=& 0, \,\, \mu=2,3,4.
\eeqa
Then the inequality (\ref{ineq}) immediately leads to,
\beqa
C(\rh(t))  &\leq&  e^{-\int^t_0(F_R(s)+G_R(s))ds}C(\rh(0)),
\label{final}
\eeqa
which establishes the first result mentioned. It is not difficult
to show that the upper bound is the minimal upper bound. To treat the
disentanglement process in general and more completely requires a
discussion of
the asymptotic behavior of the functions $F_R(s)$ and $G_R(s)$,
which is beyond the scope of the present paper.

In what follows we develop the second result mentioned above. We show
that within the general result there are very unusual and striking
specific
consequences. One example shows that, within the general exponential
character evident in (\ref{final}),  disentanglement can be completed in
a finite time while the local decoherences need an infinite time. Let us
assume that the initial density matrix is only partially coherent, but
include an arbitrary degree of non-local coherence of a familiar type
(one of the atoms is excited, but it is not certain which one). This is
easily expressed in the following form \cite{cav}
\beq
\label{initial}
\rho= \frac{1}{3}
\begin{pmatrix}
a(t) & 0 & 0 & 0 \\
0 &  b(t) &  z(t) & 0\\
0 &  z^*(t) & c(t) & 0\\
0 & 0 &  0 & d(t)
\end{pmatrix},
\eeq
where the factor 1/3 is for notational convenience. The concurrence
for this density matrix is $ C = \frac{2}{3}\max\{0,\,\, |z|-\sqrt{ad}\}$.
For simplicity, we consider an important class of mixed states with a
single
parameter $a$ satisfying initially $a\ge 0,\ d=1-a$, and $b=c=z=1$, so
then
initially $C(\rh(0))=\frac{2}{3}[1-\sqrt{a(1-a)}]$.
For $t \ge 0$ the
matrix elements are given by \beqa
a(t) &=& \ga^2_A\ga^2_Ba,\\
b(t)&=& \ga_A^2 +\ga_A^2\om^2_B a, \\
c(t)&=& \ga_B^2 +\om^2_A\ga^2_B a, \\
d(t) &=& 1-a +\om_A^2+\om_B^2+\om^2_A\om_B^2 a, \\
z(t)&=& \ga_A\ga_B. \eeqa
To simplify the calculations, we use the Markov limit results and
assume the cavities are similar so
$\ga_A = \ga_B =\ga=\exp[-\Gamma t/2],$  and $\om_A=\om_B = \om =
\sqrt{1-\exp[-\Gamma t]}.$
The concurrence for the density matrix $\rho(t)$ is given by
\beq \label{conc}
C(\rho(t)) = \frac{2}{3}\max\{0,
\ga^2 f(t)\}
\eeq
where $f(t)=1-\sqrt{a\left(1-a+2\om^2+\om^4 a\right)}.$
The sufficient condition for the concurrence
(\ref{conc}) to be zero is
\beq
1-a(1-a + 2\om^2 + \om^4a)\leq 0.
\eeq
The simplest case is $a=1$ and from this we can easily show the surprising
result that the density matrix (\ref{initial}) has a finite
disentanglement time.
That is, $C(\rho(t)) \equiv 0$ for all $t \ge t_d$, where $t_d$ is very
finite:
\beq
t_d \equiv \frac{1}{\Gamma}\ln\left[\frac{2+\sqrt{2}}{2}\right].
\eeq

\begin{figure}[t]
\includegraphics[width=6.5 cm]{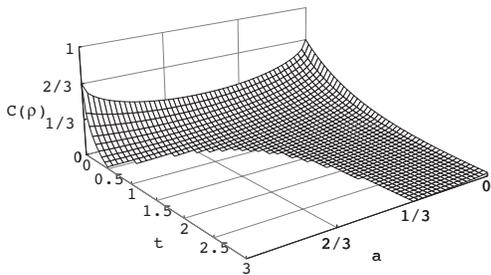}
\caption{\label{fig2} This figure shows the entanglement decay via
spontaneous emission starting  from the initially entangled states
(\ref{initial}) with $a$ between zero and 1, $d=1-a$ and $b=c=z=1$.
Finite-time complete disentanglement takes place for $a > 1/3$,
and when $a\le 1/3$ disentanglement of the
initial state is only completed asymptotically.}
\end{figure}

In response to this surprising result, a natural question will be: does
spontaneous emission cause all initially entangled two-qubit states to
disentangle
at some finite critical time? The answer is no. To see this, we consider
the entire range of different $a$ values, and plot the concurrence decay
in Fig. \ref{fig2}. The figure shows that for all $a$ values between
1/3 and 1 concurrence decay is completed in a finite time, but for smaller
$a$'s the time for complete decay is infinite.
The different behaviors exhibited over the allowed range of
$a$ values in Fig. \ref{fig2} show that our two-level atom model behaves
qualitatively differently from the continuous variable two-atom model
discussed
in \cite{hal} by Dodd and Halliwell. In that case all initially
entangled states become separable after a finite time (see also
\cite{rus}).

To summarize, we have shown for the physically fundamental and
unavoidable process of spontaneous emission that nonlocal
disentanglement times are shorter than local decoherence times for
arbitrary entangled states (pure or mixed). We based our results on
perhaps the simplest realistic decoherence scenario in which two
entangled qubits individually interact with vacuum noise. The model
allows an exact analysis and also shows, remarkably, that complete
disentanglement can be reached after only a finite time, whereas
more familiar local decoherence processes take an infinite time
to be complete. We believe our results are of generic nature.
Undoubtedly a deep understanding of the relation between decoherence
and disentanglement will be of importance for both the foundation of
quantum mechanics and practical quantum information applications.



We acknowledge constructive correspondence with J. Halliwell and
K. Wodkiewicz, and assistance with Fig. 2 from C. Broadbent. We
have obtained financial support from NSF Grants PHY-9415582
and PHY-0072359 and important assistance from L.J. Wang of
the NEC Research Institute.

\bibliography{apssamp}

\begin{thebibliography}{10}
\bibitem{com0} The derivation and memory effects
depicted by $F_R(t)$ and $G_R(t)$ of the master equation
will be discussed elsewhere. In this paper, the use of the master
equation is limited to the Markov approximation, that is, the
correlation functions are $\delta$ functions:
$\alpha(t-s)=\Gamma_A\delta(t-s)$ and $\beta(t-s)=\Gamma_B\delta(t-s)$.
\bibitem{com}In this paper, we always assume $F_R(t)$ and $G_R(t)$
to be positive functions. If $F_R(t), G_R(t)$ take negative values,
$K^{\dag}$ in equation (\ref{sum}) must be replaced with the transpose
matrix $K^T$.
\end{thebibliography}

\end{document}